\documentclass[twocolumn,letterpaper]{article}

\usepackage{anysize}
\usepackage{fancyhdr}
 
\marginsize{1.75cm}{1.75cm}{0.5cm}{0.5cm}
 \usepackage[inkscapelatex=false]{svg}
 
\usepackage[square, numbers]{natbib}
\usepackage[utf8]{inputenc} 
\usepackage[T1]{fontenc}    
\usepackage{hyperref}
\usepackage{url}            
\usepackage{booktabs}       
\usepackage{amsfonts}       
\usepackage{nicefrac}       
\usepackage{microtype}      
\usepackage{amsthm}
\usepackage{caption}
\usepackage{subcaption}
\usepackage{graphics} 
\usepackage{epsfig} 
\usepackage{times} 
\usepackage{amsmath} 
\usepackage{amssymb}  
\usepackage{makeidx}
\usepackage{float}
\usepackage{mathrsfs}
\usepackage{enumerate}
\usepackage{xr} 
\usepackage{xcolor}

\colorlet{siaminlinkcolor}{green!50!black}
\colorlet{siamexlinkcolor}{red!50!black}

\usepackage{cleveref} 

\usepackage{soul}
\usepackage{braket}
\usepackage[compat=0.3]{yquant}
\useyquantlanguage{groups}
\usetikzlibrary{quantikz2}
\usepackage{multirow}
\usepackage{scalerel}
\usepackage{siunitx}
\newcommand{\ketbra}[2]{\ket{#1}\!\bra{#2}}

\usepackage{balance} 

\usepackage{xcolor}

\begin{document}

\title{
	\vspace{-1.5cm}\rule{\linewidth}{4pt}\vspace{0.3cm} \Large \textbf{
    Controllable Quantum Memory Capacity in Quantum Reservoir \\ Networks with Tunable partial-SWAPs
	}\\ \rule{\linewidth}{1.5pt}}
	\author{Erik L. Connerty\textsuperscript{1}\thanks{Corresponding author. Email: \href{erikc@cec.sc.edu}{erikc@cec.sc.edu}}, 
     Ethan N. Evans\textsuperscript{2}\\ 
    \vspace{-.0cm}
	\small{\textsuperscript{1}University of South Carolina - Columbia, Columbia, SC, USA} \vspace{-0.0cm} \\ \small{\textsuperscript{2}Qodex Quantum, Chicago, IL, USA} \vspace{-0.0cm} \\
	}
        \date{\vspace{-.5cm}}
	\maketitle

\begin{abstract}
In the field of quantum reservoir computing (QRC), many different computational models and architectures have been proposed. From these models, we identify \textit{feedback-based} models -- which use a feedback mechanism to re-embed classical measurements from the QRC -- and \textit{recurrent models} -- which use a multi-register approach with \textit{memory} and \textit{readout} qubits -- as the two major competing architectures that have been discussed and validated on hardware. In this paper, we advance upon the recurrent architectures, which employ a two register approach to endow the QRC with a \textit{fading memory}. While these approaches have been validated on hardware and have demonstrated great real-world performance on noisy-intermediate-scale-quantum (NISQ) quantum processing units (QPUs), the exact mechanism through which the memory capacity arises is not completely understood or fully controllable. With this, we augment the recurrent approaches and present a hardware-realizable mechanism, which we call a \textit{tunable partial-SWAP}, that allows for the direct control of the rate of memory dissipation from a QRN implemented on a gate-based QPU. The theory behind this mechanism is discussed in terms of a controlled \textit{amplitude-damping channel} and validation experiments using a randomized short-term memory capacity (STMC) recall benchmark and the NARMA-5 dataset are conducted using simulation and IBM QPUs, respectively.
\end{abstract}

\section{Introduction}
QRC presents a promising framework for the near-term real-world applicability of quantum computing. As error rates for current NISQ hardware continue to decrease and QPU coupling maps become more flexible, it is pertinent to develop usable quantum machine learning (QML) algorithms that can scale to many qubits. Significant work has already been done in the pursuit of scalable quantum reservoir networks (QRNs), but it has not been firmly decided which approaches are best for differing problems and all the limitations that can be faced when deploying QRNs to quantum hardware\cite{mujal_opportunities_2021, krisnanda_experimental_2025, kornjača2024largescalequantumreservoirlearning, Hu2024,ahmed10.1098/rspa.2025.0550, Murauer_2025,feedback_PRXQuantum.5.040325,fullmeas_PhysRevResearch.6.013051,pfefferPhysRevResearch.4.033176,Connerty2026,Mujal2023,yasuda2023quantumreservoircomputingrepeated,monomi2025feedbackenhancedquantumreservoircomputing,feed_Zhu2025,feed_Paparelle2026-qj}.

The two main competing architectures for QRNs, feedback-based models \cite{ahmed10.1098/rspa.2025.0550, Murauer_2025,feedback_PRXQuantum.5.040325,fullmeas_PhysRevResearch.6.013051,pfefferPhysRevResearch.4.033176,gonon2026feedbackdrivenrecurrentquantumneural,monomi2025feedbackenhancedquantumreservoircomputing,feed_Zhu2025,feed_Paparelle2026-qj} and recurrent models \cite{Hu2024,Connerty2026}, propose different philosophies for how a QRN should retain a \textit{fading memory}. Feedback-based models use a re-encoding unitary at the beginning of each computational step to re-encode past measurements back into the quantum state, while recurrent models use a \textit{partial measure-and-reset} \cite{Hu2024} mechanism that keeps a separate register for \textit{memory} qubits and \textit{readout} qubits. Although both of these mechanisms are effective at inducing fading memory in QRNs, they each come with their own nuances and pitfalls. 

In the case of feedback-based models, fully measuring the entire quantum register completely collapses the quantum state into a classical one, while the re-encoding block adds circuit depth and further complexity at each step of computation. This full collapse also removes one of the main benefits of quantum computing: exponentially large Hilbert spaces \cite{hilb_PhysRevResearch.4.033007,schuld_supervised_2021}. When taking measurements on every qubit, irreversible memory loss is unavoidable, possibly hindering memory recall. Although quantum state tomography techniques with abundant measurements could alleviate this in theory \cite{tomog_Ghosh_2021,tomog_PhysRevX.11.041062,tomog_angelatos_PhysRevX.13.041020}, these methods would require an even higher depth unitary for re-encoding, making the technique intractable for re-encoding anything other than partial information. However, feedback-based models benefit greatly from their interpretable design, with the strength of the technique being the tunable feedback mechanism, which is easily controlled through a single hyperparameter.

In contrast, recurrent methods allow for the preservation of some superposition and entanglement through the use of a \textit{memory} register that is not measured. This technique was explored in great detail in \cite{Hu2024,Connerty2026} and has shown state-of-the-art (SOTA) performance on gate-based hardware for time-series prediction. While these recurrent methods have proven to be effective, their underlying mechanisms are still not fully understood. More specifically, in the implementation presented in both approaches, there is no single parameter that controls the memory capacity of the system. Thus, the entire circuit is subject to random weight initializations, and the mechanism through which the memory capacity is generated is not fully controllable.

With this in mind, we propose a new formulation of the recurrent models mentioned above. In particular, we determine that the underlying mechanism behind these circuits' fading memory is a \textit{controlled amplitude damping} channel that is implemented through a mechanism similar to the one shown in \cite[Fig. 8.13]{nielsen} and \cite{amplitudedamp}. We generalize this amplitude damping circuit into a tunable partial-SWAP -- a parameterized sub-circuit that partially swaps the readout and memory registers of our QRN right before measurement -- creating a weak measurement that also dampens, and thus purifies, the memory register in the process. This idea of leveraging controlled dissipation channels as a computational resource in open quantum systems has been explored in \cite{Sannia2024dissipationas,sasaki2025hamiltoniandrivenarchitecturesnonmarkovianquantum,Sannia2026}, but a clear circuit model for gate-based quantum hardware had not been established. Our method, by contrast, is implementable on any gate-based hardware with mid-circuit measurements and state-preparation operations. Additionally, our method effectively makes the embedding independent of the memory exchange processes in the implementations associated with \cite{Hu2024,Connerty2026}, giving them more design interpretability. While the work in \cite{Ricci2026} does explore a similar idea, we note that our design does not require $\mathcal{O}(N)$ depth as more qubits are added and also allows for the preservation of entanglement after each measurement, possibly increasing memory capacity as more qubits are added to the system. By using the tunable partial-SWAP operation, we gain direct control over the QRN's memory capacity and obtain something analogous to a \textit{leak rate} parameter from the classical reservoir’s leaky integration, giving a clear quantum analogue to the classical echo-state network (ESN) \cite{jaeger2001}.


\section{Methods}
\begin{figure*}[t]
    \centering
    \includegraphics[width=\linewidth, keepaspectratio]{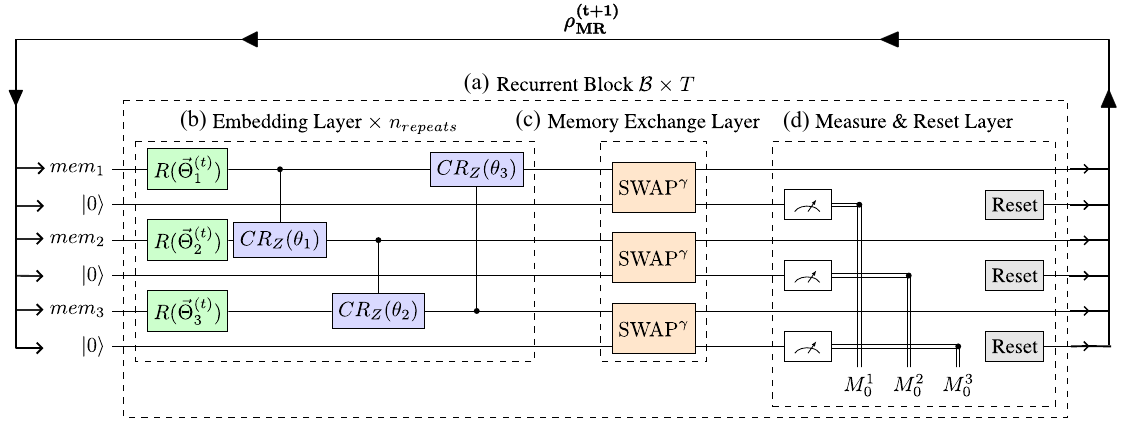}
    \vspace{1pt}
    \caption{Proposed recurrent partial-SWAP QRN. The circuit is composed of two registers: \textit{memory} and \textit{readout}, labeled $mem$ and $\ket{0}$, respectively. a) The Recurrent Block, $\mathcal{B}$, is the main outer-block and runs recurrently over the length of the time-series input $T$. Three subcircuits perform the information processing for the partial-SWAP QRN at each time-step. b) The Embedding Layer, input data enters in through the $R(\vec{\Theta}_i)$ gate onto the $i$th memory qubit through the mapping detailed in Eq.~\eqref{eq:input-embedding}. A CRZ entangling layer parameterized by $\theta_i$ is performed at the end of each application of the reuploading scheme. c) The Memory Exchange Layer performs the partial-SWAP operation parameterized by the hyperparameter $\gamma$, moving some of the information of the \textit{memory} register into the \textit{readout} register. d) The Measure \& Reset Layer, where the \textit{readout} register is measured in the Pauli-Z basis, and then subsequently reset back to $\ket{0}_R$ for the next time-step of computation. Putting these layers together creates a controlled amplitude-damping channel that purifies and weakly-measures the \textit{memory} register.}
    \label{fig:QRN_Circuit}
\end{figure*}

In this section, we provide a detailed breakdown of each layer of the partial-SWAP QRN. The circuit can be seen in Figure~\ref{fig:QRN_Circuit}. We note that the initial state for this algorithm is $\rho^{(0)}_{MR} = \ketbra{0}{0}^{\bigotimes n_{qubits}}$, where $n_{qubits}$ is the total number of qubits used in the system for both the \textit{memory} and \textit{readout} registers.

\subsection{Embedding Layer}
The embedding layer, which handles the input encoding, is a fully customizable circuit component that can theoretically be replaced with any embedding scheme. We describe our implementation here.
\paragraph{Context Window}
First, a context window of the classical data is mapped to rotation angles through a scheme similar to \cite{Connerty2026}. Let $u_t$ be the scalar input time-series at time-step $t$ where $t\in \set{1,2,\dots,T}$ and $T$ is the total length of the time-series data. The input to the QRN at time $t$ is then $\vec{x}_t = (u_{t},u_{t-1},\dots,u_{t-c})$, where $c$ is the context length for the encoding. We encode the data at each $t$ onto $\frac{n_{qubits}}{2}$ memory qubits, indexed by $j$, with three Euler angles per memory qubit indexed by $k$. The input layer and bias weight tensors are given as $W^{in} \in \mathbb{R}^{c \times \frac{n_{qubits}}{2} \times 3}$ and $W^{bias} \in \mathbb{R}^{\frac{n_{qubits}}{2} \times 3}$, respectively. The hidden layer weights for the two-qubit gates are indicated as $\vec{w}^{hidden} \in \mathbb{R}^{\frac{n_{qubits}}{2}}$. All weights are sampled from a uniform distribution in the interval $(0,\pi]$. This gives the input embedding rotations $\Theta^{(t)} \in \mathbb{R}^{\frac{n_{qubits}}{2} \times 3}$ as a linear function of the current input.

\begin{align}\label{eq:input-embedding}
\Theta_{j,k}^{(t)} = \sum_{i=1}^{c} \vec{x}_{t-i+1} \, W^{in}_{i,j,k} + W^{bias}_{j,k},
\end{align}
For example, the entry $\Theta_{1,3}^{(t)}$ refers to the 3rd Euler angle for the 1st memory qubit in the circuit at time $t$. We assign the parameter $\theta_i = w^{hidden}_i$ as the hidden layer weight for each qubit's two-qubit gate at every time-step. These weights are used to dictate the rotations on each qubit at each time-step.

\paragraph{Data Reuploading Scheme}
Afterwards, a data reuploading scheme is used to add nonlinearity to the circuit \cite{P_rez_Salinas_2020,chu2022qmlperrortolerantnonlinearquantum,Govia_2022,schuld_supervised_2021,express_Xiong2025,express_schutte2025expressivitylimitsquantumreservoir}.
The computed angles from Eq.~\eqref{eq:input-embedding} then construct the gate
\begin{equation}\
R^{(t)}_j(\Theta^{(t)}_{j1}, \Theta^{(t)}_{j2}, \Theta^{(t)}_{j3}) = R_x(\Theta^{(t)}_{j1}) R_y(\Theta^{(t)}_{j2}) R_x(\Theta^{(t)}_{j3}),
\label{eq:rotation}
\end{equation}
where $R_x$ and $R_y$ are arbitrary single-qubit rotation gates along their respective Pauli basis, $\Theta^{(t)}$ is the input-dependent Euler angle rotation matrix, and $R_j^{(t)}$ is the $t$th rotation gate for the $j$th memory qubit. With the generated angles, we construct the embedding unitary
\[
U^{(t)}_{\mathrm{emb}}(\Theta^{(t)},\vec{\theta})
\;:=\;
\bigl(U_{\mathrm{CRZ}}(\vec{\theta})\,U_{\mathrm{rot}}\Theta^{(t)})^{n_{\mathrm{repeats}}} \otimes \mathbb{I}_R
\]
where $U^{(t)}_{\mathrm{emb}}$ is the time-dependent unitary acting only on the \textit{memory} register, $\vec{\theta}$ are the entangling weights for the CRZ gate, and $n_{repeats}$ is the number of times to apply the reuploading scheme. This unitary evolves the system
\begin{equation}
\rho^{\prime(t)}_{MR} = U_{\mathrm{emb}}^{(t)}\bigl(\Theta^{(t)},\vec{\theta}\bigr)\, \rho^{(t)}_{MR} \, U_{\mathrm{emb}}^{(t)\dagger}\bigl(\Theta^{(t)},\vec{\theta}\bigr)
\end{equation}
at each time-step. 

\subsection{Memory Exchange Layer}\label{subsec:pswap}
\begin{figure}[ht]
    \centering
    \resizebox{\linewidth}{!}{









\begin{tikzpicture}
   \begin{yquantgroup}
      \registers{
         qubit {} q[2];
      }
      \circuit{
         cnot q[1]| q[0];
         box {$\text{\footnotesize $X$}^\gamma$} q[0] | q[1];
         cnot q[1] | q[0];
      }
      \equals
      \circuit{
        box {$\mathrm{SWAP}^{\gamma}$} (q);
      }
   \end{yquantgroup}
\end{tikzpicture}}
    \vspace{1pt}
    \caption{The partial-SWAP circuit. A single parameter $\gamma \in (0,1]$ is used to control the strength of the memory dampening effect. This is implemented as a matrix exponential in both cases.}
    \label{fig:partial-SWAP}
\end{figure}
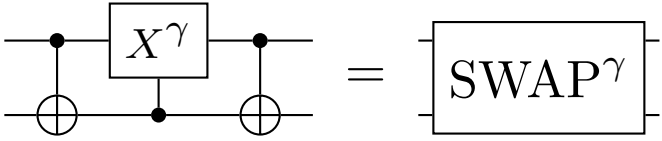

We implement a subcircuit called a partial-SWAP between adjacent memory and readout qubits in the Memory Exchange Layer. This important subcircuit block can be shown to act as a controlled \textit{amplitude-damping channel} when the \textit{readout} qubit is always reset to the eigenstate $\ket{0}$. This idea is elaborated on further in Supplementary Note 1: Amplitude Damping Channel with partial-SWAP \ref{supp1:amplitude}. With the use of this mechanism, the aforementioned Embedding Layer becomes fully customizable, while the Memory Exchange Layer gains the important hyperparameter  $\gamma \in (0,1]$ that functions similarly to a \textit{leak rate} from classical reservoir networks. More specifically, at $\gamma = 1$, the partial-SWAP is simply a canonical SWAP operation that reads all the information off the coupled memory qubit while returning it to state $\ket{0}$. However, when set in the range $0 < \gamma < 1$, the partial-SWAP only reads partial-information and can leave some residual information on the coupled memory qubit. We hypothesize that optimal partial-SWAP parameters will always be somewhere in the range $0 < \gamma < 1$ for memory dependent tasks, as extreme values will read either too little or too much information to have reliable predictive performance or memory capacity, respectively. A diagram of two equivalent partial-SWAP circuits is shown in Figure \ref{fig:partial-SWAP}. This layer evolves the system
\begin{equation}
\rho^{\prime\prime(t)}_{MR} = U_{SWAP}(\gamma)\rho^{\prime(t)}_{MR} \, U^{\dagger}_{SWAP}(\gamma),
\end{equation}
where $U_{SWAP}(\gamma)$ is the partial-SWAP unitary given in Supplementary Note 1 Eq.~\eqref{eq:partial-SWAP}.

\subsection{Measure \& Reset Layer}
In the final layer of the partial-SWAP QRN, measurements are taken as counts, and sampled bitstrings, i.e. (`00',`01','10',`11'), are reconstructed into an empirical distribution with discrete probabilities at each time-step $t$. These probabilities are placed into a feature matrix
\begin{equation}\label{eq:features}
A \in \mathbb{R}^{T \times F},\qquad A=
\begin{pmatrix}
p_{1,1} & p_{1,2} & \dots & \dots\\
p_{2,1} & p_{2,2} & \dots & \dots\\
\vdots & \vdots & \ddots & \vdots\\
\vdots & \vdots & \dots & p_{T,F}
\end{pmatrix}
\end{equation}
where $T$ is the total number of time-steps in the input time-series, $F$ is the dimensionality of the Hilbert space on the readout register. The entries of the feature matrix are given by
\[
p_{t,i} = \frac{O_{t,i}}{n_{shots}}
\]
where $O$ is the raw observed count for the $i$th bitstring, and $n_{shots}$ is the number of measurement shots used in the experiment. The probabilities in each row satisfy $\sum_{i=0}^{F}p_{t,i} \approx 1$ up to some numerical precision. Lastly, the reset channel is applied to the \textit{readout} register, making the final updated state
\begin{equation}
\rho^{(t+1)}_{MR}
=
\operatorname{Tr}_R\!\left[\rho^{\prime\prime (t)}_{MR}\right]
\otimes
\ketbra{0}{0}_R
\end{equation}
for the next time-step of computation.

\section{Results}
We now describe the results obtained with the partial-SWAP QRN in both simulation and hardware experiments. Simulations were conducted using the noiseless Aer simulator \cite{qiskit2024}, while hardware results were obtained using the \texttt{ibm\_boston} QPU.

\subsection{Simulation: Short-Term Memory Capacity} \label{subsubsec:stmc}
Similarly to the feedback-based models studied in \cite{Murauer_2025,monomi2025feedbackenhancedquantumreservoircomputing}, we quantify the QRN’s ability to retain recent inputs with a standard short-term memory capacity (STMC)
task. The network is driven by an i.i.d. scalar input sequence
$u_t \sim Uniform(0,1)$, and the resulting sampled reservoir states
are recorded at each time step. Stacking these row-wise yields the feature matrix in Eq.~\eqref{eq:features}. For each delay $\tau \in \{0,\dots,-10\}$, features and targets are aligned so that the feature vector at time step $t$ is paired with the target
\[
\hat{y}_t^{(\tau)} = u_{t+\tau}. 
\]

After discarding the first $n_{\mathrm{washout}}=15$ aligned samples, we split the remaining data into $n_{\mathrm{train}}=700$ training samples and $n_{\mathrm{test}}=275$ test samples. A linear model is then trained to learn the delayed reconstruction map for each delay $\tau$
\[
y_t^{(\tau)} = \vec{w}_\tau^{\top}\vec{a}_t + b_\tau.
\]
In matrix form, the model weights are obtained by minimizing
\begin{equation}\label{eq:ridge}
\mathcal{L}(\vec{w}_\tau)
=
\left\|A_{\mathrm{train}}\vec{w}_\tau - \mathbf{\hat{y}}_{\mathrm{train}}^{(\tau)}\right\|_2^2
+
\alpha \|\vec{w}_\tau\|_2^2,
\end{equation}
with ridge parameter $\alpha = \num{1e-5}$ \cite{hoerl1970ridge}. The trained weights are then applied to $A_{\mathrm{test}}$ and compared to $y_{\mathrm{test}}^{(\tau)}$.

\begin{figure*}[t!]
    \centering
    \includegraphics[width=\linewidth, keepaspectratio]{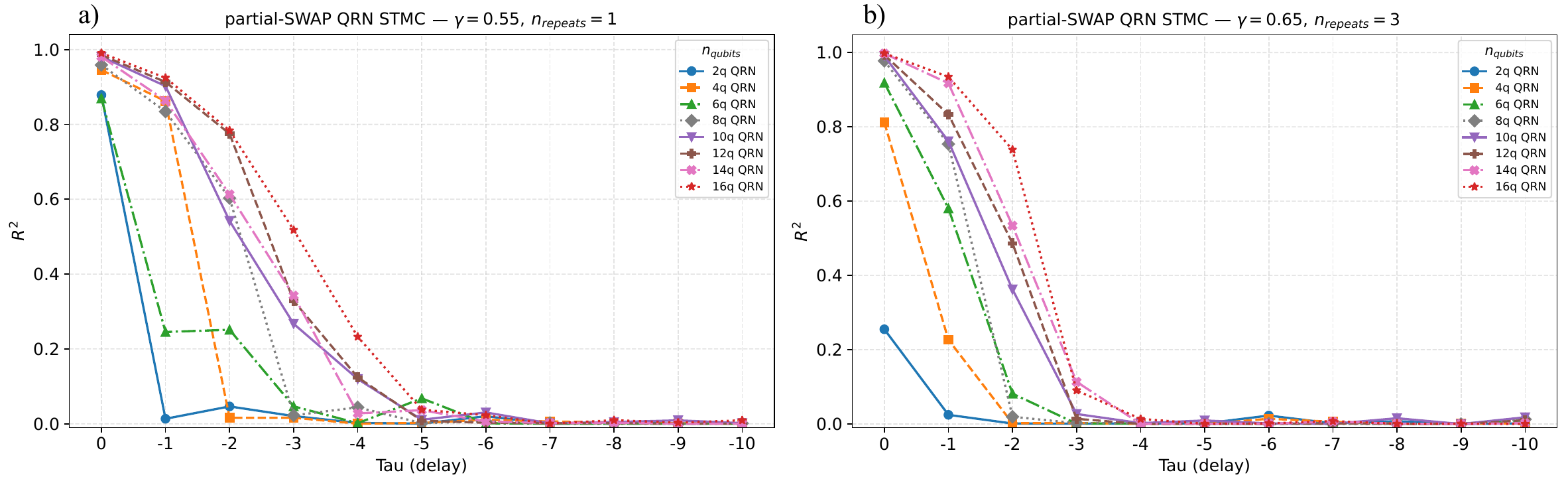}
    \vspace{1pt}
    \caption{STMC Best $R^2$ Comparison for two Configurations of the QRN. Plots depicting the coefficient of determination ($R^2$) values for differing $n_{qubits}$ values. \textit{Higher is better}. A context length of $c = 1$ is used in this experiment, giving a worst-case demonstration of memory-recall.  The best partial-SWAP strength $\gamma$ was chosen from each $n_{repeats}$ configuration for the $n_{qubits} = 16$ circuit in Figure \ref{fig:STMC_gamma} to give a more detailed look at memory performance. a) Depicts the partial-SWAP QRN with $n_{repeats} = 1$ reuploading blocks at partial-SWAP strength $\gamma = .55$. b) Depicts the partial-SWAP QRN with $n_{repeats} = 3$ reuploading blocks at $\gamma = .65$. As shown in Figure \ref{fig:STMC_gamma}, the performance on the STMC task was generally better in the $n_{repeats} = 1$ configuration.}
    \label{fig:STMC_tau}
\end{figure*}

Memory performance at each delay is measured on the test set using the coefficient of determination
\begin{equation}
R_\tau^2 = 
\frac{
\mathrm{cov}\!\left(\hat{y}_{\mathrm{test}}^{(\tau)}, y_{\mathrm{test}}^{(\tau)}\right)^2
}{
\sigma_{\hat{y}_{\mathrm{test}}^{(\tau)}}^2 \sigma_{y_{\mathrm{test}}^{(\tau)}}^2
},
\end{equation}
which is the standard delay-dependent memory score. Figure \ref{fig:STMC_tau} shows the best-performing values of $\gamma$ for different embedding-layer configurations with varying $n_{repeats}$.

To compare all partial-SWAP configurations, we also summarize short-horizon reconstruction over delays $\tau = 0,\dots,-4$ using the root mean squared error (RMSE) on the test set,
and report the mean short-horizon error
\begin{equation}
\overline{\mathrm{RMSE}}_{0:4}
=
\frac{1}{5}\sum_{\tau=0}^{4}\mathrm{RMSE}_\tau.
\end{equation}
We chose this cutoff because the NARMA-5 task in Section \ref{subsub:narma} also requires a fading memory of 5 time-steps.

\begin{figure*}[t]
    \centering
    \includegraphics[width=\linewidth, keepaspectratio]{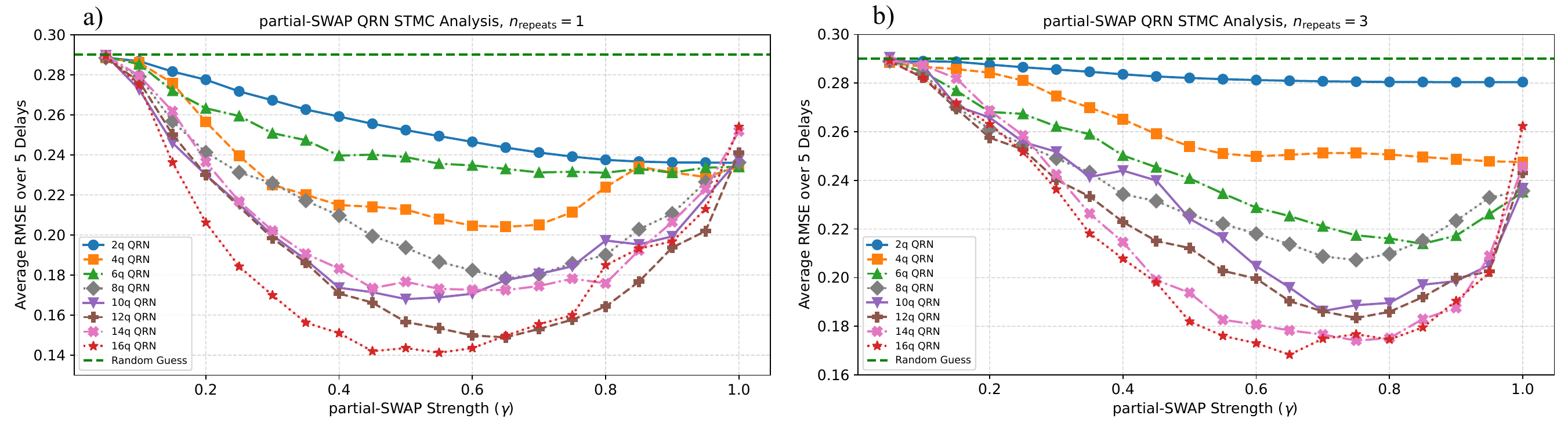}
    \vspace{1pt}
    \caption{STMC partial-SWAP Strength Comparison. An ablation study over various parameters of partial-SWAP strength $\gamma$ and $n_{qubits}$ against average RMSE over 5 time-delays $\tau$ on the STMC task. \textit{Lower is better}. A context length of $c = 1$ is used in this experiment, giving a worst-case demonstration of memory-recall. a) Depicts the partial-SWAP QRN with $n_{repeats} = 1$ reuploading blocks. b) Depicts the partial-SWAP QRN with $n_{repeats} = 3$ reuploading blocks. As hypothesized, optimal partial-SWAP strength parameters lie in the range $0 < \gamma < 1$, and performance deteriorates when getting too close to either extreme. Overall, the best performance on this task is given by the $n_{repeats} = 1$ configuration, which is believed to be due in part to the reduced nonlinearity of this circuit. In both cases, we see a strong effect of the partial-SWAP strength parameter $\gamma$, which when varied drastically affects the outcome of the experiment. The random guess is given as $\sqrt{\mathrm{Var(U(0,1))}}$}
    \label{fig:STMC_gamma}
\end{figure*}

As discussed in Section \ref{subsec:pswap}, if the partial-SWAP strength controls memory capacity, then an optimal value should occur at some $0 < \gamma < 1$, corresponding to the lowest RMSE. Because $\gamma=1$ is equivalent to a full SWAP, memory should degrade as $\gamma$ approaches one. Likewise, for $\gamma$ near zero, too little information is transferred from the \textit{memory} register to support accurate prediction. We observe strong evidence for this hypothesis in nearly all tested partial-SWAP QRN configurations in Figure \ref{fig:STMC_gamma}. The main deviations likely arise because only a single seed was tested per configuration and because some lower-qubit circuits have limited memory capacity to begin with. With more samples, we expect the convex trends seen in some larger circuits to become typical. Owing to the computational cost of simulating these large circuits with shot-based simulators across many configurations, we report results only for seed $\eta = 42$.

Overall, these results show that the hyperparameter $\gamma$ directly affects QRN memory performance, and that STMC performance generally improves with increasing qubit counts -- something that has not been shown with feedback-based models. We also note that the partial-SWAP QRN uses a context length of only $c=1$ in these experiments, representing a worst-case setting for memory recall. These results therefore provide useful guidance for hyperparameter selection when tuning the QRN for a given task, as discussed further in the next section. The lower RMSE observed for $n_{repeats}=1$ is likely due to the additional nonlinearity introduced at larger $n_{repeats}$, which is detrimental to a purely memory-based task such as the STMC task. All experimental parameters are listed in Table \ref{tab:stmc_params}, and additional results are provided in Supplementary Note 3: Supplementary Results \ref{supp:supplementary_results}.

\subsection{Simulation: NARMA-5}\label{subsub:narma}

To evaluate the QRN's nonlinear temporal processing, we consider the fifth-order nonlinear autoregressive moving average (NARMA-5) task, a standard reservoir computing benchmark. For an input sequence $z_t \sim \mathrm{Uniform}(0,0.5)$, the target is generated recursively as
\begin{equation}
\hat{y}_{t+1}
=
0.3 \hat{y}_t
+
0.05 \hat{y}_t \sum_{i=0}^{4} \hat{y}_{t-i}
+
1.5 z_{t-4} z_t
+
0.1,
\end{equation}
with the first 5 samples discarded to allow system startup.

Each experiment uses $n_{\mathrm{total}}=1000$ input samples and the corresponding NARMA-5 targets. As in the STMC task, the QRN is driven by the input signal $z_t$, and the readout at each time-step is converted into a feature vector, as shown by Eq.~\eqref{eq:features}. The resulting feature representation $A$ is then used to learn the one-step-ahead mapping
\[
\vec{w}_\tau^{\top}\vec{a}_t + b_\tau = \hat{y}_{t+1}.
\]

We use $n_{\mathrm{train}}=750$ training samples and $n_{\mathrm{test}}=250$ test samples, with a washout of $n_{\mathrm{washout}}=15$ removed from the training data and labels. The model weights are obtained via ridge regression, similarly to Eq.~\eqref{eq:ridge}.
The trained model is then evaluated with $A_{\mathrm{test}}$ against $\hat{\mathbf{\vec{y}}}_{\mathrm{test}}$.

\begin{figure*}[t!]
    \centering
    \includegraphics[width=\linewidth, keepaspectratio]{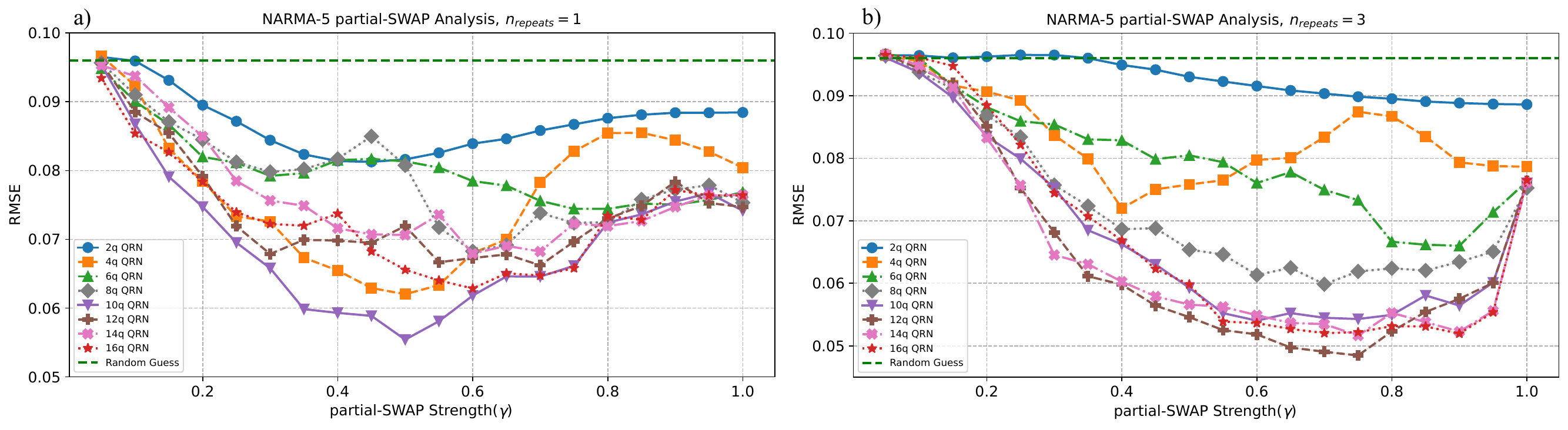}
    \vspace{1pt}
    \caption{NARMA-5 Comparison for Various Configurations of partial-SWAP QRN. An ablation study over various parameters of partial-SWAP strength $\gamma$, $n_{qubits}$, and $n_{repeats}$ against RMSE. \textit{Lower is better}. a) Denotes the task run with $n_{repeats} = 1$ reuploading blocks. b) Denotes the task run with $n_{repeats} = 3$ reuploading blocks. As predicted by the results in Figure \ref{fig:STMC_gamma}, the partial-SWAP strength tends to have a clear optimal point for each configuration. Because of the need for increased nonlinearity in the NARMA-5 task, we see improved performance with $n_{repeats} = 3$ reuploading blocks. This result demonstrates that memory capacity is almost certainly dictated by the partial-SWAP strength $\gamma$, with extreme values that tend towards full SWAP or identity operations performing poorly. The random guess is given as $\sqrt{\mathrm{Var(\hat{y})}}$.}
    \label{fig:NARMA_aggregate}
\end{figure*}

Simulation of the circuit is performed to obtain an ablation over multiple different partial-SWAP $\gamma$ configurations applied to the NARMA-5 task. Figure \ref{fig:NARMA_aggregate} depicts the result of this experiment for $n_{repeats} \in \set{1,3}$ circuits. As hinted before, the NARMA-5 task is in strong agreement with the STMC task of \ref{subsubsec:stmc}: we see that the optimal partial-SWAP strength $\gamma$ for the STMC task is very close to the best value for the NARMA-5 task. We also see that the addition of more reuploading blocks improves performance on this task, in contrast to Figure \ref{fig:STMC_gamma}, likely due to the increased need for nonlinearity. Consequently, there does seem to be an unknown limitation that causes stagnation in performance past $n_{qubits} = 12$. We leave this observation up for further discussion. 

\begin{figure*}[t!]
    \centering
    \includegraphics[width=\linewidth, keepaspectratio]{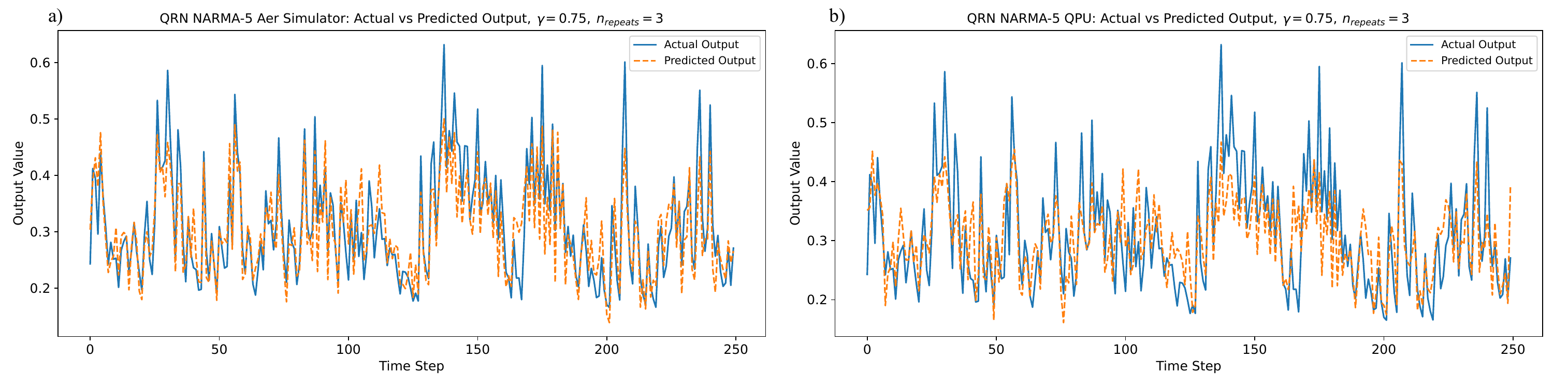}
    \vspace{1pt}
    \caption{NARMA-5 Results for both the Aer Simulator and IBM QPU using the partial-SWAP QRN. The NARMA-5 task test predictions for both the noiseless Aer simulator (a), and the \texttt{ibm\_boston} QPU (b). For both experiments, $n_{qubits} = 12$ qubits were used and the number of reuploading blocks in the embedding layer was $n_{repeats} = 3$. The RMSE for a) .0484 and b) .0646 show strong agreeance between the noiseless implementation and the QPU implementation.}
    \label{fig:NARMA_sim_qpu}
\end{figure*}

A specific prediction of the NARMA-5 system for the highest performing QRN is shown in Figure \ref{fig:NARMA_sim_qpu} (a). The RMSE for this run was $.0484$. Overall, the NARMA-5 task proved to be an informative demonstration for the partial-SWAP QRN and further validates the claim that the partial-SWAP hyperparameter $\gamma$ is indeed a predictor of memory capacity and thus the performance of the QRN. We list all the parameters for the simulated NARMA-5 task in Table \ref{tab:narma_aer_params} and give additional results for this experiment in Supplementary Note 3: Supplementary Results \ref{supp:supplementary_results}.

\subsection{Simulation: Comparison to ESN}\label{subsub:classical}
\begin{figure}[ht!]
    \centering
    \includegraphics[width=\linewidth, keepaspectratio]{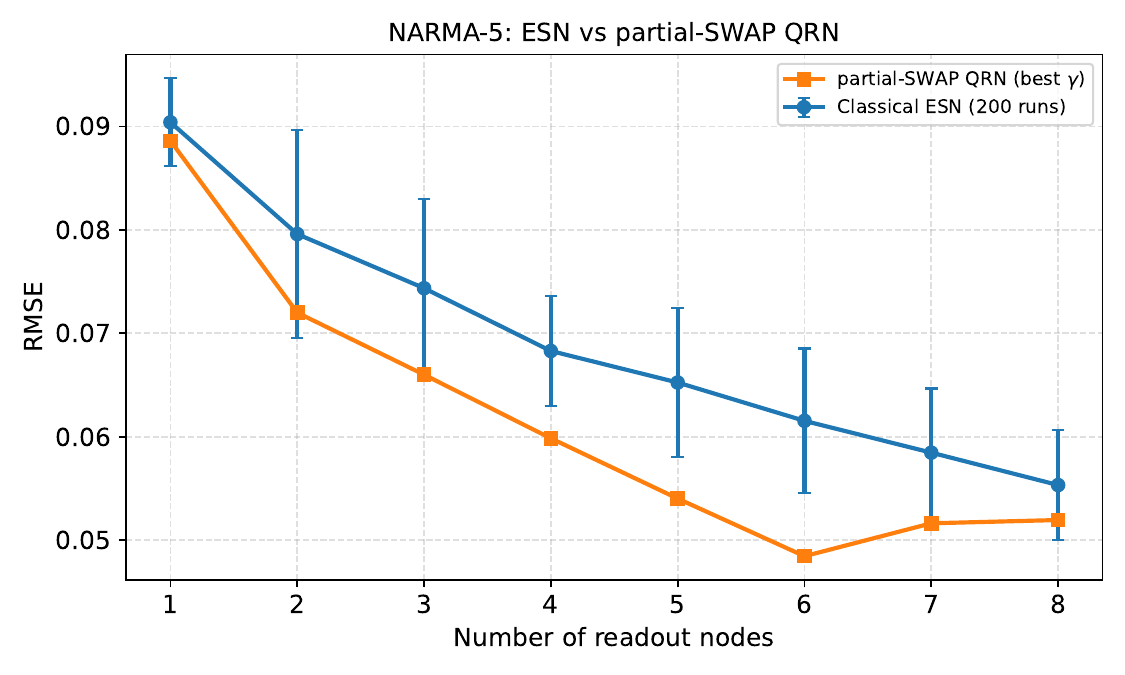}
    \vspace{1pt}
    \caption{partial-SWAP QRN vs Classical ESN on NARMA-5 Task. A comparison of the partial-SWAP QRN to a classical ESN. \textit{Lower is better}. The ESN is run 200 times on the NARMA-5 task with different seeds to generate error bars. The number of readout nodes is set to be the same size as the QRN's \textit{readout} register.}
    \label{fig:NARMA_classical_compare}
\end{figure}

We provide a brief comparison of the partial-SWAP QRN to a classical ESN \cite{jaeger2001} similar to the experiment done in \cite{Connerty2026}. This experiment uses a standard ESN with parameters

\begin{equation}
     N_{nodes} \in \set{1,2,3,4,5,6,7,8}, \rho(W) = .9, \lambda = .5
\end{equation}

where $N_{nodes}$ is the number of nodes in the reservoir, $\rho(W)$ is the spectral radius of the hidden layer weight matrix, and $\lambda$ is the leak rate, which controls the rate at which information is ``leaked'' from the system. The nonlinear activation was chosen as $\tanh{h_t}$, where $h_t$ is the $t$th hidden state of the ESN. Comparisons are made using 200 runs of the ESN with different seeds. For fairness, the ESN is chosen to have the same number of nodes as the QRN's readout register size. This is done because the QRN only effectively processes information using the \textit{memory} register, with the readout serving as a way to read partial information. Thus, the comparison is done with $\frac{n_{qubits}}{2} = N_{nodes}$ in each case. Figure \ref{fig:NARMA_classical_compare} shows this comparison. Although we are only able to use one sample of the partial-SWAP QRN due to the complexities and resource intensiveness of large quantum circuit simulations, there is an apparent edge for the quantum model in just about every case.

\subsection{Hardware: NARMA-5} \label{subsec:qpu}
Motivated by the strong results obtained with the noiseless Aer simulator, we ran a proof-of-concept for tunable partial-SWAP QRNs on the \texttt{ibm\_boston} QPU. This experiment tested the theory of the tunable partial-SWAP on present-day NISQ hardware. Similar to the experiment described in \cite{Connerty2026}, we used a deterministic delay inserted into the circuit after each measurement to allow for longer circuits with more measurements to run without any ``buffer overflow'' errors \cite{Hu2024}. While this was done as a precautionary measure due to past troubles, it seems that the issue may have been alleviated to some extent, as this error was not observed a single time during experimentation. Still, out of caution, the experiments were run using an $8\mu$s delay at each time-step of computation. Additionally, the target $n_{shots} = 60,000$ shots for the experiment are broken into 30 jobs of 2,000 shots each.

After compiling the circuit for the target \texttt{ibm\_boston} QPU using $n_{qubits} = 12$ qubits, a circuit of $c_{depth} =$ 203,133 is obtained using the same $n_{total} = 1000$ data points from the NARMA-5 task. This qubit count was chosen because it achieved the best result in the noiseless Aer simulator and was also expected to have a lower circuit depth than other configurations. Since our QRN algorithm uses output counts instead of expectation values, the number of shots needed to recover the counts grows rapidly with more qubits. With this configuration, the execution time for one shot of the circuit was no less than $8,000\mu$s -- well over the $\mathcal{T}1$ and $\mathcal{T}2$ times of the \texttt{ibm\_boston} QPU. The reason this circuit is believed to be able to execute for this long recurrently is due to the partial-SWAP's amplitude-damping effect, which also serves as an error-mitigation method for long term execution. It can be shown that the purity of the reduced density matrix over the \textit{memory} register $\rho_{M}$ tends to $1$ through the partial-SWAP and measure-and-reset layer by gradually returning that register to the pure state $\ket{0}^{\bigotimes \frac{n_{qubits}}{2}}$ over time. We examine this idea further in Supplementary Note 2: Purifying Effect of the partial-SWAP Channel \ref{supp2:purity}.

The final result of the hardware experiment is depicted in Figure \ref{fig:NARMA_sim_qpu} (b). This experiment resulted in an RMSE value of $.0646$, which is ~$33$\% higher than the noiseless Aer simulator value of $.0484$. Despite this difference in performance — which is to be expected in the NISQ era — we see strong agreement with theoretical results and real-world performance. It is clear that the QRN was still able to function in the high-noise environment on the NARMA-5 task and that the partial-SWAP mechanism, in conjunction with the measure-and-reset layer, worked to endow the QRN with a fading memory as well as mitigate errors over time. The full parameters for the IBM QPU experiment can be found in Table \ref{tab:narma_qpu_params}.

\section{Discussion}
We presented a recurrent QRN with a controllable memory-capacity endowed through a mechanism called a partial-SWAP. This mechanism allows for the introduction of a single tunable hyperparameter $\gamma$ for directly controlling the memory-capacity of a QRN. The STMC experiments demonstrated definitively that the QRN memory capacity was controllable using the partial-SWAP strength parameter $\gamma$. These experiments also showed that the memory-capacity of our circuit increased with the number of qubits in the system, a property that has not been demonstrated in feedback-based models to the authors' best knowledge. With higher qubit QPUs becoming available, this sort of result looks promising for the scalability of QRNs with respect to memory capacity.

In the NARMA-5 task, we showed that real-world applicability is possible with the proposed tunable partial-SWAP QRN. By demonstrating that the QRN could learn the dynamics of the highly nonlinear and memory dependent NARMA-5 system, we show that all necessary components of a functional QRN are present in our model. We also demonstrated a slight hint of a quantum advantage in Section \ref{subsub:classical}. While there were some concerns about the scalability of the model due to the apparent bottleneck at $n_{qubits} = 12$ qubits in Figure \ref{fig:NARMA_aggregate} (b), we note that this work was mainly concerned with examining the effect of the partial-SWAP mechanism, and that future work could improve this through even higher qubit counts or a better embedding scheme, potentially.

Lastly, we ran a high-impact experiment using current SOTA IBM hardware, bringing all of the theory together into a NISQ era demonstration of time-series prediction. This experiment demonstrated that the partial-SWAP QRN was not only tolerant to noise due to the discussed purification process in Section \ref{subsec:qpu}, but it could also retain its nonlinear information processing capabilities when deployed on NISQ era hardware. With a circuit composed of $203,133$ gates and a time-series of $1000$ data points, we demonstrated that current gate-based NISQ hardware was capable of highly-nonlinear temporal information processing.

\section{Conclusions}
The results in this paper definitively demonstrate the efficacy and controllability offered by the tunable partial-SWAP subcircuit when used in conjunction with the measure-and-reset schemes in \cite{Hu2024,Connerty2026}. With this advancement in theory, it should be possible to build more robust and interpretable QRNs, while bringing QRN algorithms closer to the controllability of their classical counterparts. As more and more QPUs with lower error rates and better connectivity come online, we foresee an explosion of interest in QML. Thus, it is pertinent to get ahead of the curve and develop the necessary techniques and algorithms for creating truly quantum models that still have some of the nice characteristics their classical counterparts possess. This paper pushes the needle further towards the development of truly controllable quantum recurrent neural networks (QRNNs), in the opinion of the authors. Tunable partial-SWAPs are a trivial way to create fading memory in a quantum circuit and only require $\mathcal{O}(N)$ qubits asymptotically. This method can be used in a plethora of ways, and the authors hope to see it developed further in the future as more QML models are created.

\section{Acknowledgements}
We acknowledge the use of IBM Quantum Credits via the IBM Quantum Startups Program for this work, the computational resources provided by the Theia high performance computing cluster at the University of South Carolina which is supported by National Science Foundation Grant No. 2320292, and the technical assistance and resources provided by Research Computing at the University of South Carolina (RRID:SCR\_027488). The views expressed in this paper are those of the authors and do not reflect the official policy or position of any of the aforementioned parties. The author ELC acknowledges support from the DoW SMART (Science, Mathematics, and Research for Transformation) Scholarship.


\begin{table}[ht!]
        \centering
        \scriptsize 
        \begin{tabular}{c|c c}
        Name & Parameter & Value \\
        \hline
        Total \# of Qubits & $n_{qubits}$ & $\set{2,4,6,8,10,12,14,16}$ \\
        \hline
        partial-SWAP Strength & $\gamma$ & $\set{.05,.1,.15,\dots,1.0}$ \\
        \hline
        \# of Shots & $n_{shots}$ & 30,000 \\
        \hline
        \# of Re-uploading Blocks & $n_{repeats}$ & $\set{1,3}$ \\
        \hline
        \# of Total Data Points & $n_{total}$ & 1000 \\
        \hline 
        \# of Training Data Points & $n_{train}$ & 700 \\
        \hline 
        \# of Test Data Points & $n_{train}$ & 275 \\
        \hline
        \# of Washout Points & $n_{washout}$ & 15 \\
        \hline 
        Context Length & $c$ & 1 \\
        \hline 
        Regularization Parameter & $\alpha$ & $\num{1e-5}$ \\
        \hline 
        Seed & $\eta$ & 42 \\
        \hline
        Delays & $\tau$ & $\set{0,-1,-2,...,-10}$ \\
        \end{tabular}
        \caption{STMC Aer Simulator Parameters}
        \label{tab:stmc_params}
\end{table}

\begin{table}[ht!]
        \centering        
        \scriptsize 
        \begin{tabular}{c|c c}
        Name & Parameter & Value \\
        \hline
        Total \# of Qubits & $n_{qubits}$ & $\set{2,4,6,8,10,12,14,16}$ \\
        \hline
        partial-SWAP Strength & $\gamma$ & $\set{.05,.1,.15,\dots,1.0}$ \\
        \hline
        \# of Shots & $n_{shots}$ & 60,000 \\
        \hline
        \# of Re-uploading Blocks & $n_{repeats}$ & $3$ \\
         \hline
        \# of Total Data Points & $n_{total}$ & 1000 \\
        \hline
        \# of Training Data Points & $n_{train}$ & 750 \\
        \hline
        \# of Test Data Points & $n_{test}$ & 250 \\
        \hline
        \# of Washout Points & $n_{washout}$ & 15 \\
        \hline 
        Context Length & $c$ & 5 \\
        \hline 
        Regularization Parameter & $\alpha$ & $\num{1e-4}$ \\
        \hline 
        Seed & $\eta$ & 42 \\
        \end{tabular}
        \caption{NARMA-5 Aer Simulator Parameters}
        \label{tab:narma_aer_params}
\end{table}

\begin{table}[ht!]
        \centering
        \small 
        \begin{tabular}{c|c c}
        Name & Parameter & Value \\
        \hline
        Total \# of Qubits & $n_{qubits}$ & $12$ \\
        \hline
        partial-SWAP Strength & $\gamma$ & $.75$ \\
        \hline
        Total \# of Shots & $n_{shots}$ & 60,000 \\
        \hline
        \# of Shots per job & n/a & 2,000 \\
        \hline
        Total \# of jobs & n/a & 30 \\
        \hline
        \# of Re-uploading Blocks & $n_{repeats}$ & $3$ \\
         \hline
        \# of Total Data Points & $n_{total}$ & 1000 \\
        \hline
        \# of Training Data Points & $n_{train}$ & 750 \\
        \hline
        \# of Test Data Points & $n_{test}$ & 250 \\
        \hline
        \# of Washout Points & $n_{washout}$ & 15 \\
        \hline 
        Context Length & $c$ & 5 \\
        \hline 
        Regularization Parameter & $\alpha$ & $\num{1e-4}$ \\
        \hline 
        Seed & $\eta$ & 42 \\
        \hline
        Circuit Depth & $c_{depth}$ & 203,133
        \end{tabular}
        \caption{NARMA-5 IBM QPU Parameters}
        \label{tab:narma_qpu_params}
\end{table}

\bibliography{sn-bibliography}
\balance

\clearpage
\appendix
\onecolumn

\appendix

\section{Supplementary Note 1: Amplitude Damping Channel with partial-SWAP}\label{supp1:amplitude}

Here we show that the partial-SWAP with measure-and-reset creates an amplitude damping channel. This important property is what allows the partial-SWAP circuit to have a finite \textit{fading memory} and provides access to partial information transfer to the readout qubit while necessarily damping the memory qubit to a fixed-point. Let the memory qubit be \(M\) and the readout qubit be \(R\), with the
readout initialized in \(\ket{0}_R\). Consider the two-qubit partial-SWAP unitary
\[
U_{SWAP}(\gamma)
=
\operatorname{CNOT}_{M\to R}\;
C_R(X_M^\gamma)\;
\operatorname{CNOT}_{M\to R} = \mathrm{SWAP}^{\gamma}
\]
where \(C_R(X_M^\gamma)\) denotes a controlled-\(X^\gamma\) gate acting on the
memory qubit, conditioned on the readout qubit, and $\gamma \in (0,1]$. Since \(X^2=I\), the
fractional power \(X^\gamma\) may be written as
\begin{equation}\label{eq:Xdecomp}
    X^\gamma = A I + B X
\end{equation}
where,
\[
\qquad
A=\frac{1+e^{i\pi\gamma}}{2} = e^{i\frac{\pi\gamma}{2}}\cos{(\frac{\pi\gamma}{2})},
\qquad
B=\frac{1-e^{i\pi\gamma}}{2} = -ie^{i\frac{\pi\gamma}{2}}\sin{(\frac{\pi\gamma}{2})}
\]
Therefore
\[
|A|^2=\cos^2\!\left(\frac{\pi\gamma}{2}\right),
\qquad
|B|^2=\sin^2\!\left(\frac{\pi\gamma}{2}\right).
\]
Now define
\begin{equation}\label{eq:damping_prob}
p := |B|^2 = \sin^2\!\left(\frac{\pi\gamma}{2}\right), 
\qquad
1-p := |A|^2 = \cos^2\!\left(\frac{\pi\gamma}{2}\right),
\end{equation}
so \(p\) will be the damping probability as per Eq.~\eqref{eq:Xdecomp}.
The final partial-SWAP matrix is
\begin{equation}\label{eq:partial-SWAP}
U_{SWAP}(\gamma)=
\begin{pmatrix}
1 & 0 & 0 & 0\\
0 & \multicolumn{2}{c}{\multirow{2}{*}{$\vcenter{\hbox{\hspace{0.5em}\LARGE $X^\gamma$}}$}} & 0\\
0 & \multicolumn{2}{c}{} & 0\\
0 & 0 & 0 & 1
\end{pmatrix}
=
\begin{pmatrix}
1 & 0 & 0 & 0\\
0 & A & B & 0\\
0 & B & A & 0\\
0 & 0 & 0 & 1
\end{pmatrix}.
\end{equation}

With this definition, we apply the partial-SWAP operation to the input state at a given time-step. Let the input state of the memory qubit at time be $t$, where $t \in [1,2,\dots,T]$ be
\[
\rho^{(t)}_M=
\begin{pmatrix}
\rho^{(t)}_{00} & \rho^{(t)}_{01}\\
\rho^{(t)}_{10} & \rho^{(t)}_{11}
\end{pmatrix}.
\]
Since the readout qubit is always reset at the end of each time-step, it is always in the state \(\ket{0}\) before applying the partial-SWAP, so the joint input state before the partial-SWAP layer is
\[
\rho_{MR}^{(t)}=\rho^{(t)}_M\otimes \ket{0}\!\bra{0}_R
=
\begin{pmatrix}
\rho^{(t)}_{00} & 0 & \rho^{(t)}_{01} & 0\\
0 & 0 & 0 & 0\\
\rho^{(t)}_{10} & 0 & \rho^{(t)}_{11} & 0\\
0 & 0 & 0 & 0
\end{pmatrix}.
\]
The output state is generated by evolving $\rho^{(t)}_{MR}$ unitarily
\[
\rho^{\prime(t)}_{MR}=U_{SWAP}(\gamma) \rho_{MR}^{(t)} U_{SWAP}(\gamma)^\dagger =
\begin{pmatrix}
\rho^{(t)}_{00} & \rho^{(t)}_{01}B^* & \rho^{(t)}_{01}A^* & 0\\
\rho^{(t)}_{10}B & \rho^{(t)}_{11}|B|^2 & \rho^{(t)}_{11}BA^* & 0\\
\rho^{(t)}_{10}A & \rho^{(t)}_{11}AB^* & \rho^{(t)}_{11}|A|^2 & 0\\
0 & 0 & 0 & 0
\end{pmatrix}.
\]

Finally, taking the partial trace of $R$
\[
\rho^{\prime(t)}_M=\operatorname{Tr}_R(\rho_{MR}^{\prime(t)})
=
    \begin{pmatrix}
        \rho^{(t)}_{00}+|B|^2\rho^{(t)}_{11} & A^*\rho^{(t)}_{01}\\
        A\rho^{(t)}_{10} & |A|^2\rho^{(t)}_{11}
    \end{pmatrix}
 =
    \begin{pmatrix}
        \rho^{(t)}_{00}+p\rho^{(t)}_{11} & A^*\rho^{(t)}_{01}\\
        A\rho^{(t)}_{10} & (1-p)\rho^{(t)}_{11}
    \end{pmatrix}.
\]
Thus population initially in \(\ket{1}_M\) is transferred to \(\ket{0}_M\)
with probability \(p\). 
Since \(|A|=\sqrt{1-p}\), the partial-SWAP implements amplitude damping with a damping parameter \(p\), up to a phase difference in the coherence terms
\[
A=e^{i\phi}\sqrt{1-p},
\qquad
\phi=\frac{\pi\gamma}{2},
\]
and the final state is
\begin{equation}\label{eq:final_amplitude}
\rho_M^{\prime(t)}=
\begin{pmatrix}
\rho_{00}^{(t)}+p\rho_{11}^{(t)} & e^{-i\phi}\sqrt{1-p}\,\rho_{01}^{(t)}\\
e^{i\phi}\sqrt{1-p}\,\rho_{10}^{(t)} & (1-p)\rho_{11}^{(t)}
\end{pmatrix}.
\end{equation}
Therefore, the partial-SWAP circuit realizes amplitude
damping on the memory qubit with damping parameter
\(p=\sin^2(\pi\gamma/2)\). This derivation can be easily generalized to arbitrary memory and readout register sizes as long as they have an equal number of qubits.

\section{Supplementary Note 2: Purifying Effect of the partial-SWAP Channel}\label{supp2:purity}

Following along the same lines as Supplementary Note~\ref{supp1:amplitude}, we show that the partial-SWAP circuit ultimately purifies the memory register through a simple example. From Eq.~\eqref{eq:final_amplitude}, we see the partial-SWAP circuit induces an amplitude damping channel on the memory qubit \(M\), with the damping parameter given by Eq.~\eqref{eq:damping_prob}. Then for an input state
\[
\rho_M^{(t)}=
\begin{pmatrix}
\rho_{00}^{(t)} & \rho_{01}^{(t)}\\
\rho_{10}^{(t)} & \rho_{11}^{(t)}
\end{pmatrix},
\]
we can show that a single application of the partial-SWAP gives
\[
\rho_M^{\prime(t)}
=
\begin{pmatrix}
\rho_{00}^{(t)}+p\rho_{11}^{(t)} & A^*\rho_{01}^{(t)}\\
A\rho_{10}^{(t)} & (1-p)\rho_{11}^{(t)}
\end{pmatrix},
\qquad |A|^2=1-p.
\]

Under repeated partial-SWAPs plus measure-and-resets, the population in \(\ket{11}_{MR}\) evolves as
\[
\rho_{11}^{(t+n)}=(1-p)^n\rho_{11}^{(t)},
\]
where $n$ is the number of applications of the channel. The off-diagonal terms decay as
\[
\rho_{01}^{(t+n)}=(A^*)^n\rho_{01}^{(t)},
\qquad
\rho_{10}^{(t+n)}=A^n\rho_{10}^{(t)}.
\]
Using the fact that the trace must sum to one, the \(\ket{00}_{MR}\) population is therefore
\[
\rho_{00}^{(t+n)}=1-\rho_{11}^{(t+n)}=1-(1-p)^n\rho_{11}^{(t)}.
\]
After \(n\) repeated applications, the final state is
\begin{equation}\label{eq:purity_repeated}
\rho_M^{(t+n)}
=
    \begin{pmatrix}
    1-(1-p)^n\rho_{11}^{(t)} & (A^*)^n\rho_{01}^{(t)}\\
    A^n\rho_{10}^{(t)} & (1-p)^n\rho_{11}^{(t)}
    \end{pmatrix}.
\end{equation}

For any nontrivial partial-SWAP, \(0<p\le 1\), since \(|A|=\sqrt{1-p}<1\), it follows that
\[
(1-p)^n\rho_{11}^{(t)}\to 0,
\qquad
(A^*)^n\rho_{01}^{(t)}\to 0,
\qquad
A^n\rho_{10}^{(t)}\to 0
\qquad \text{as } n\to\infty.
\]
Thus, all terms decay to zero except the $\ket{00}_{MR}$ population
\[
\lim_{n\to\infty}\rho_M^{(t+n)}
=
\begin{pmatrix}
1 & 0\\
0 & 0
\end{pmatrix}
=
\ket{0}\!\bra{0}.
\]

Since \(\ket{0}\!\bra{0}\) is a pure state, its purity is
\[
\operatorname{Tr}\!\left[(\ket{0}\!\bra{0})^2\right]=1.
\]
This demonstrates that the repeated application of the partial-SWAP channel drives the memory register toward the pure ground state \(\ket{0}\!\bra{0}\). In this sense, the partial-SWAP has a purifying effect on the memory qubit, allowing for indefinite operation as long as some memory loss is tolerated by the problem. We do note that while our experiments use the parameter $\gamma \in (0,1]$, it can be shown through the relationship in Eq.~\eqref{eq:damping_prob} that $p$ is also $\in (0,1]$ for this interval of $\gamma$.

\section{Supplementary Note 3: Supplementary Results}\label{supp:supplementary_results}
\begin{figure*}
    \centering
    \caption{STMC partial-SWAP Strength Comparison for $n_{repeats = 6}$}
    \includegraphics[width=\linewidth, keepaspectratio]{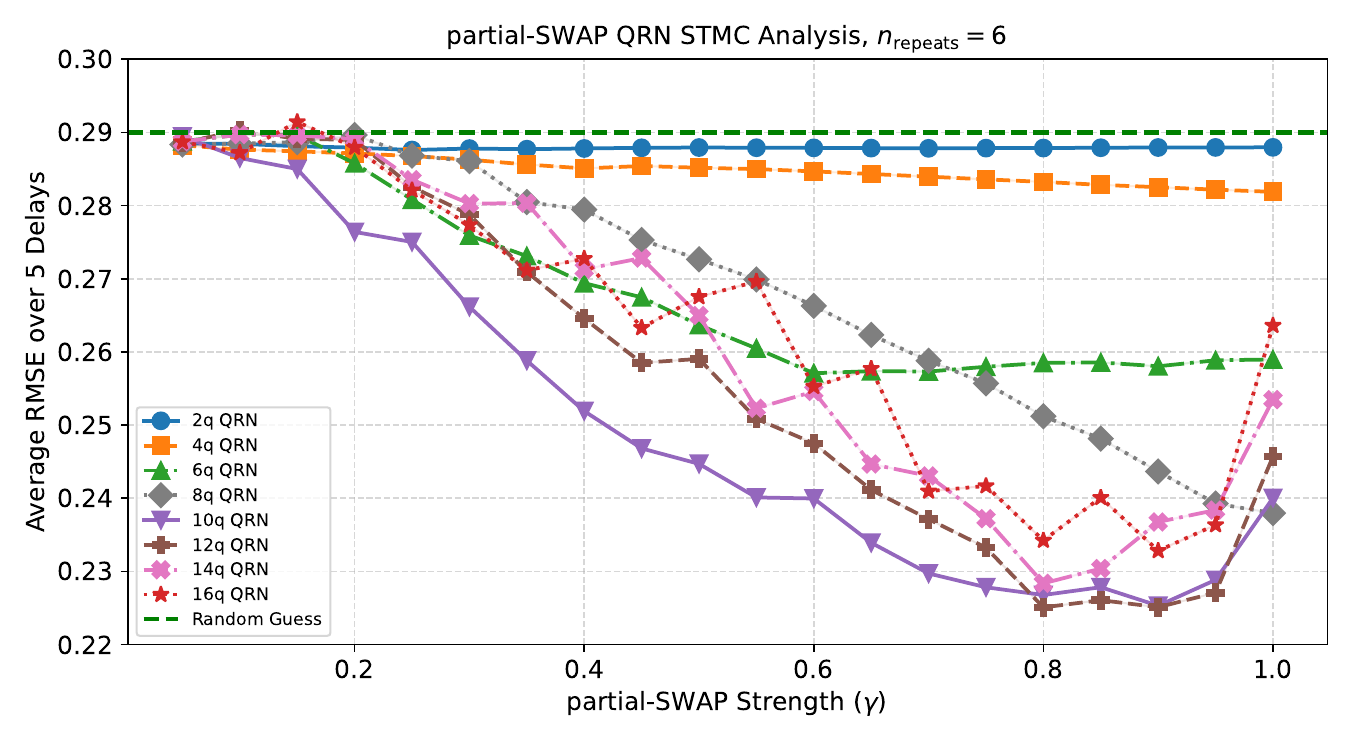}
    \vspace{1pt}
    \caption*{STMC results for $n_{repeats} = 6$. Performance is believed to be lower than in the lower repeat block experiements of Figure \ref{fig:STMC_gamma} due to the higher nonlinearity added by more repeat blocks. The random guess is given as $\sqrt{\mathrm{Var(U(0,1))}}$}
    \label{fig:STMC_rep6}
\end{figure*}

\begin{figure*}
    \centering
    \caption{NARMA-5 partial-SWAP Strength Comparison for $n_{repeats = 6}$}
    \includegraphics[width=\linewidth, keepaspectratio]{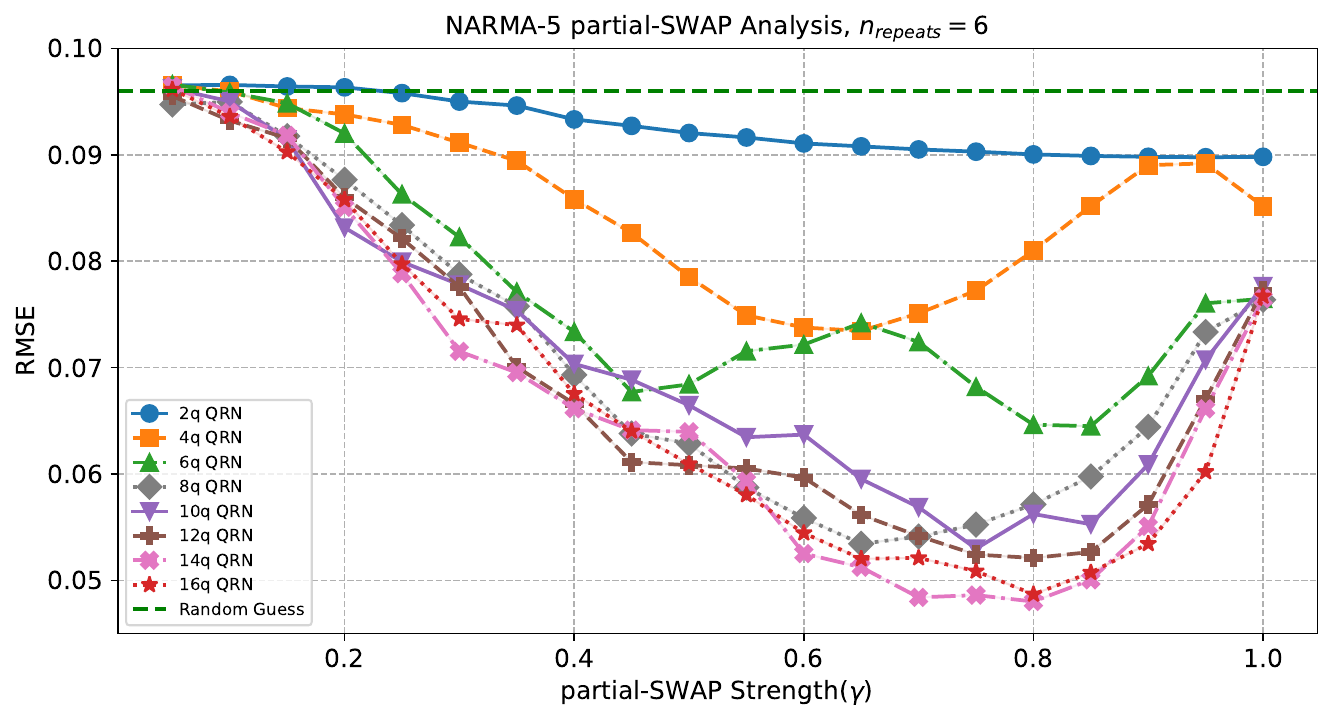}
    \vspace{1pt}
    \caption*{NARMA-5 partial-SWAP strength results for $n_{repeats = 6}$. While performance remains good in this configuration, more repeat blocks add higher circuit depth, and the tradeoff was not seen as worthwhile given the incremental performance improvement over the configuration in Figure \ref{fig:NARMA_aggregate}. The reduced performance seen with even numbered $n_{repeats}$ in \cite{Connerty2026} seems to be alleviated by the partial-SWAP mechanism, as performance remains high in both the $n_{repeats} = 3$ and $n_{repeats} = 6$ configurations. The random guess is given as $\sqrt{\mathrm{Var(y)}}$}
    \label{fig:NARMA_rep6}
\end{figure*}

Here we present results for $n_{repeats} = 6$ experiments for the STMC and NARMA-5 tasks of Sections 3.1.1 \& 3.1.2, respectively. For the STMC task, we see in Figure \ref{fig:STMC_rep6} that the performance generally decreases with more reuploading blocks. As mentioned before, this is likely due to the increased nonlinearity of the circuit hindering memory recall. For the NARMA-5 task, Figure \ref{fig:NARMA_rep6} depicts better performance than the $n_{repeats} = 3$ in some configurations, but due to the higher depth of circuits in this configuration, they were not used for deployment on the IBM QPU.

\end{document}